\documentclass[english,twocolumn,10pt,superscriptaddress,prl]{revtex4-2}
\usepackage[T1]{fontenc}
\usepackage[utf8]{inputenc}
\usepackage{color}
\usepackage{babel}
\usepackage{amsmath}
\usepackage{amssymb}
\usepackage{graphicx}
\usepackage[normalem]{ulem}
\usepackage[pdfusetitle,
bookmarks=true,bookmarksnumbered=false,bookmarksopen=false,
 breaklinks=false,pdfborder={0 0 1},backref=false,colorlinks=true]
 {hyperref}
\hypersetup{
 linkcolor=blue, citecolor=cyan}

\makeatletter
\usepackage{amsthm}

\usepackage{xcolor}

\makeatother

\begin{document}
\title{Late-Time Relaxation from Landau Singularities}
\author{Dong-Lin Wang}
\email{donglinwang@mail.ustc.edu.cn}

\affiliation{Department of Modern Physics and Anhui Center for fundamental Sciences
(Theoretical Physics), University of Science and Technology
of China, Anhui 230026, China}
\affiliation{Department of Physics, The University of Tokyo, 7-3-1 Hongo, Bunkyo-ku,
Tokyo 113-0033, Japan}
\author{Shi Pu}
\email{shipu@ustc.edu.cn}

\affiliation{Department of Modern Physics and Anhui Center for fundamental Sciences
(Theoretical Physics), University of Science and Technology
of China, Anhui 230026, China}
\affiliation{Southern Center for Nuclear-Science Theory (SCNT), Institute of Modern
Physics, Chinese Academy of Sciences, Huizhou 516000, Guangdong Province,
China}

\begin{abstract}
Nonlinear hydrodynamic interactions can change the relaxation of fluctuations from exponential to power-law decay at late times. Schwinger-Keldysh effective field theory provides a standard framework for describing such fluctuation effects, where the nonlinear late-time behavior is encoded in loop corrections. Extracting this behavior requires identifying the singularities of loop integrals, whose structure becomes increasingly intricate beyond simple models.
We apply Landau singularity analysis to two-point functions in effective field theories and determine the singularities induced by nonlinear interactions without performing the loop integrations explicitly. From these frequency-space singularities, we extract nonlinear relaxation modes that control the late-time behavior. When gapless modes are present, these modes produce power-law decay at late times. Our results give a systematic singularity-based description of nonlinear late-time relaxation in a broad class of macroscopic effective theories.
\end{abstract}

\maketitle

\paragraph{Introduction.}---
At finite temperature, microscopic thermal motion makes fluctuations an intrinsic component of macroscopic dynamics, producing variations in quantities such as the  number density, pressure, and energy density. In large systems these fluctuations are often small in relative magnitude, scaling down with increasing particle number, and were therefore historically treated as negligible corrections \citep{gibbs1902a}. 
Yet in specific physical regimes, fluctuations are not negligible corrections but can control observable phenomena. For example, in relativistic heavy-ion collisions, fluctuations of conserved charges provide probes of critical behavior in the QCD phase diagram \citep{Stephanov:1998dy,Gupta:2011wh,Asakawa:2015ybt}. In two-dimensional systems with continuous symmetry, fluctuations preclude long-range order at finite temperature, as expressed by the Mermin-Wagner theorem \citep{Mermin:1966fe,Hohenberg:1967zz,Berezinskii:1970pzv}. In gravitational-wave detectors, thermal fluctuations in mirror coatings provide an important contribution to the displacement noise that limits precision measurements \citep{Harry:2001iw, Harry:2006fi}. These examples illustrate that fluctuations are not merely background noise, but dynamical physical effects that can shape macroscopic observables.

A central question is then how fluctuations relax. While fluctuations originate from microscopic randomness, the relaxation of fluctuations encodes the dynamical and transport properties of a system. At early times, fluctuation dynamics can depend sensitively on microscopic details.  
At late times, however, fast microscopic transients are expected to decay, and the remaining relaxation is governed primarily by symmetries, conservation laws, and constitutive relations rather than by the detailed microscopic Hamiltonian \citep{Landau:1980mil}. In this  late-time regime, relaxation is controlled by  macroscopic processes, including momentum diffusion, heat conduction, and charge diffusion, whose connection to equilibrium fluctuations is formalized by the fluctuation-dissipation theorem \citep{Kubo:1966fyg}.
Since these processes control the slow macroscopic response of the system, late-time relaxation provides a sensitive probe of macroscopic dynamics. For example, it reveals critical slowing down near phase transitions \citep{Hohenberg:1977ym}, signals the emergence of hydrodynamics in quantum many-body systems \citep{Lux:2013mpl,Zu:2021irm,Joshi:2021fno}, and captures the breakdown of  ergodicity in glassy systems where memory effects prevent fluctuations from fully decaying \citep{Palmer:1982pm,Binder:1986zz}.

To characterize this late-time behavior, one must go beyond linear dissipative dynamics. In linear response theory, fluctuation eigenmodes propagate and dissipate independently, and their contribution to correlation functions typically decays exponentially \citep{Kadanoff:1963axw}. Nonlinear interactions invalidate this independent-mode picture by coupling a fluctuation to products of other modes, thereby generating singularities absent from the linear spectrum \citep{Kovtun:2012rj}. In the long-wavelength limit, these nonlinear singularities can dominate the  late-time behavior and replace purely exponential relaxation by the long-time tail, namely the power-law decay of correlation functions produced by nonlinear hydrodynamic interactions. This power-law decay was first identified in the velocity autocorrelation function \citep{Alder:1970zza,Ernst:1970asymt} and is now understood as a consequence of nonlinear coupling among slow hydrodynamic modes. In current correlators, for example, nonlinear terms involving products of slow variables lead to momentum integrals over hydrodynamic propagators and yield power-law decay at late times \citep{Arnold:1997gh,Kovtun:2003vj,Delacretaz:2020nit}.
In recent years, fluctuation effects and nonlinear hydrodynamic interactions have been formulated systematically within Schwinger-Keldysh effective field theory (SK-EFT) \citep{Haehl:2015uoc,Crossley:2015evo,Jensen:2017kzi}. In this framework, linearized classical dynamics is encoded in tree-level propagators, while nonlinear interactions and non-Gaussian fluctuations arise from higher-order terms in the effective Lagrangian. The corresponding late-time contributions to correlation functions are therefore encoded in loop integrals. These integrals have been evaluated explicitly in nonlinear diffusion models at one-loop \citep{Chen-Lin:2018kfl,Abbasi:2022aao} and two-loop \citep{Jain:2020zhu} order, and the analysis was extended to arbitrary-loop banana diagrams in Ref.~\citep{Grozdanov:2024fle}.

Although much progress has been made, existing calculations mainly establish important examples and do not yet provide a general characterization of nonlinear late-time relaxation. The main challenge is that late-time behavior is controlled by singularities of loop integrals, whose locations, types, and physical relevance are difficult to determine beyond simple models or low-loop order. As a result, the extent to which a common late-time structure emerges remains unclear. This limitation obstructs a systematic understanding of fluctuation relaxation and of fluctuation phenomena more broadly.

This motivates an approach that determines the relevant singularities directly, without performing the loop integrations explicitly. 
In this work, we apply Landau singularity analysis \citep{Landau:1959fi} to loop corrections in SK-EFT. This method was originally developed to analyze singularities of Green's functions in quantum field theory \footnote{Landau singularity analysis has recently also been applied to correlation functions in first-order hydrodynamics \citep{Abbasi:2022rum}.}. This approach identifies the decay modes generated by nonlinear interactions through simple algebraic conditions. We obtain the general form of the leading singularities in banana diagrams and show how they control late-time relaxation. When gapless modes are present, the same structure yields the power-law decay at late times. Our results provide a systematic description of nonlinear late-time relaxation and clarify how frequency-space singularities control the asymptotic behavior in time.

\vspace{0.3cm}
\paragraph{Correlation functions in effective theory.}---
We begin with the SK-EFT formalism \citep{Haehl:2015uoc,Crossley:2015evo,Jensen:2017kzi}. For dynamical fields $\varphi=\{\varphi_i\}$, the theory is formulated in terms of the $r/a$ variables $\varphi_r=\{\varphi_{ri}\}$ and $\varphi_a=\{\varphi_{ai}\}$, where $\varphi_r$ describes deviations from equilibrium and $\varphi_a$ encodes stochastic fluctuations. For a near-equilibrium system, the effective Lagrangian takes the form
\begin{equation}
\mathcal{L}_{\mathrm{eff}}=\varphi_{a}(\mathcal{R}\varphi_{r}+\mathcal{S}\varphi_{a})+\mathcal{L}_{\textrm{int}},
\label{eq:EffectiveLagrangian}
\end{equation}
with $\mathcal{R}(\partial)$ and $\mathcal{S}(\partial)$ polynomial in spacetime derivatives. 
Here $\mathcal{L}_{\textrm{int}}$ denotes the nonlinear interaction part, whose terms all contain at least one $\varphi_a$.
Terms linear in $\varphi_a$ yield the classical equation of motion without noise, while terms nonlinear in $\varphi_{a}$ encode noise effects, including the Gaussian noise term $\varphi_a \mathcal{S}\varphi_a$ \citep{Kamenev2011}.

Given Eq.~(\ref{eq:EffectiveLagrangian}), the correlation functions can be computed perturbatively. The full two-point function $G$ can be expanded perturbatively as
\begin{equation}
G\approx G^{(0)} + G^{(0)}\Sigma G^{(0)}+\dots,
\label{eq:DysonEq}
\end{equation}
where we use the superscript $(0)$ denotes leading-order quantities in this work. In particular, $G^{(0)}$ is the leading-order propagator and $\Sigma$ is the self-energy \citep{Kamenev2011}. 
In the $r/a$ basis, the retarded, advanced, and symmetrized correlators are $G_R=G_{ra}=\left\langle \varphi_{r}\varphi_{a}\right\rangle$, $G_A=G_{ar}=\left\langle \varphi_{a}\varphi_{r}\right\rangle$, and $G_S=G_{rr}=\left\langle \varphi_{r}\varphi_{r}\right\rangle$, respectively.
Since $G_A(t)=0$ for $t>0$, the late-time behavior is determined by $G_R$ and $G_S$. After Fourier transforming Eq.~(\ref{eq:DysonEq}), one finds that the late-time behavior is governed by the singularities of $G^{(0)}$ and $\Sigma$ in momentum space.

\vspace{0.3cm}
\paragraph*{Poles at leading order.}---
We next analyze the  singularities of the free two-point functions $G^{(0)}$. We focus on cases in which these singularities are \emph{simple poles} in the complex frequency plane, as in diffusion and more general hydrodynamic models \citep{Kadanoff:1963axw,Kovtun:2012rj}. Here, by a simple pole we mean a first-order pole with finite, nonzero residue. This is a key assumption in our analysis.

The poles of $G^{(0)}$ can be identified directly from the free part of $\mathcal{L}_{\mathrm{eff}}$. When they are simple, $G^{(0)}$ can be decomposed into a sum over pole contributions. For example, if
$G_{R}^{(0)}\sim[(p^{0}-\mathfrak{w}_{1})(p^{0}-\mathfrak{w}_{2})]^{-1}$,
where $\mathfrak{w}_{n}(\mathbf{p})$ are basic decay modes satisfying
$\mathrm{Im}\,\mathfrak{w}_{n}<0$ for $\mathbf{p}\neq0$, then 
by partial-fraction
decomposition \citep{ahlfors1979complex}, 
$G_{R}^{(0)}\sim(\mathfrak{w}_{1}-\mathfrak{w}_{2})^{-1}[(p^{0}-\mathfrak{w}_{1})^{-1}-(p^{0}-\mathfrak{w}_{2})^{-1}]$.
This decomposition generalizes directly to a finite number of simple poles.
More generally, 
suppose the free retarded propagator $G_{R}^{(0)}$
has $N_{0}$ poles, so that 
$G_{R}^{(0)}\propto\prod_{n=1}^{N_{0}}[p^{0}-\mathfrak{w}_{n}(\mathbf{p})]^{-1}$. Then,
\begin{equation}
G_{R}^{(0)}(p)=\sum_{n=1}^{N_0}\frac{K_n(\mathbf{p})}{p^0-\mathfrak{w}_n(\mathbf{p})},
\label{eq:GraGeneralForm0}
\end{equation}
where $K_{n}(\mathbf{p})$ depends on $\mathbf{p}$ and is determined by the form of the propagator.
The free advanced propagator $G_{A}^{(0)}$ and
the symmetrized propagator $G_{S}^{(0)}$ can be
decomposed analogously.

\begin{figure}
\begin{centering}
\includegraphics[width=0.6\linewidth]{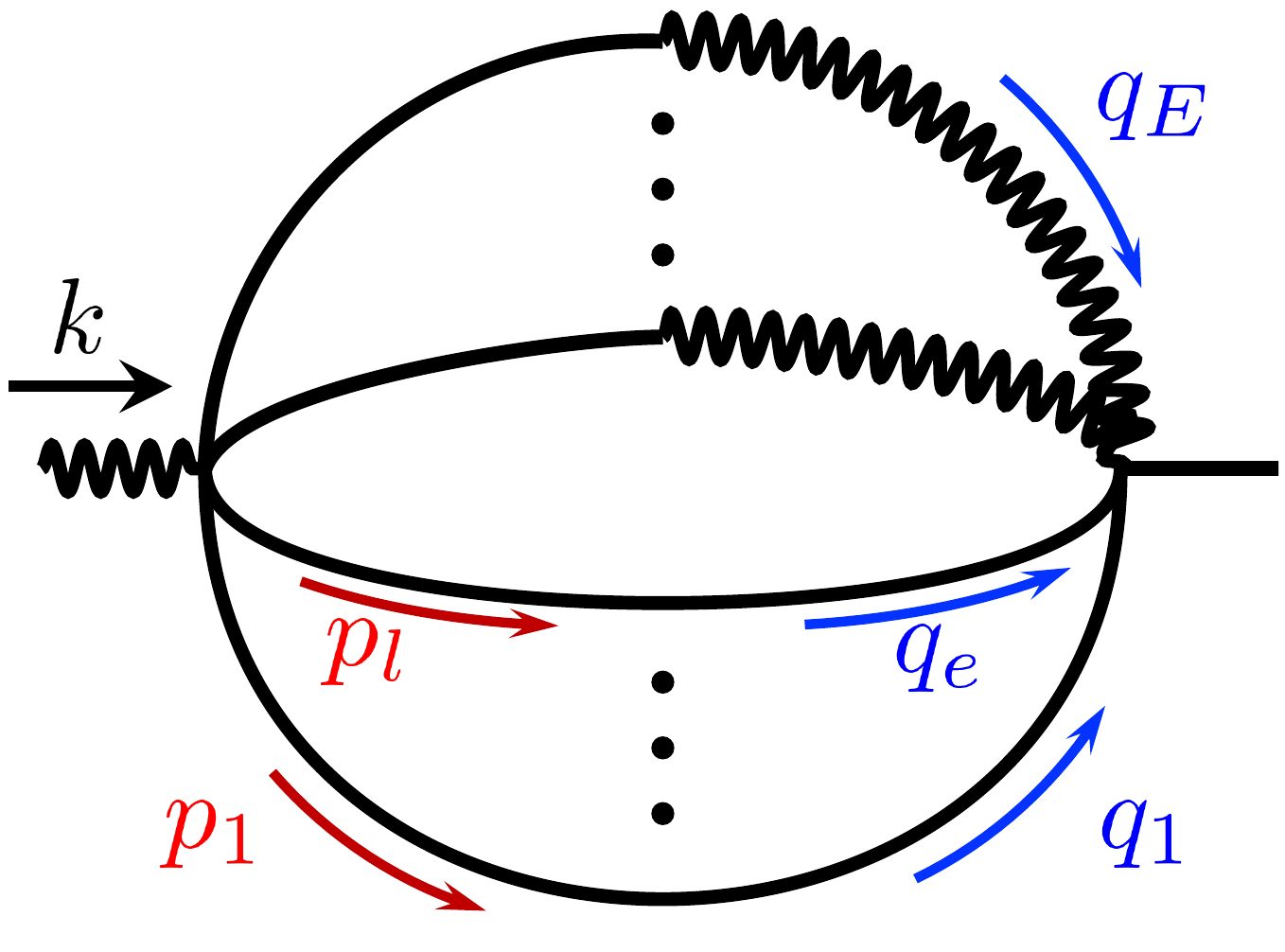} 
\par\end{centering}
\caption{
Illustration of a banana diagram. Wavy and solid lines represent the $\varphi_a$ and $\varphi_r$ legs, respectively. The external momentum is $k$; $p_l$ denote the independent loop momenta, and $q_e$ the internal momenta carried by the propagators. For illustration, we choose a simple routing in which $q_e=p_e$ for $e<E$ and  $q_E=k-\sum_{l=1}^{L} p_l$.
}\label{fig1}
\end{figure}

\vspace{0.3cm}
\paragraph*{Singularities in self-energy and Landau equations}--- 
At next-to-leading order in Eq.~\eqref{eq:DysonEq}, the full propagator involves the self-energy $\Sigma$, whose singularities arise from nontrivial loop integrals. 
Consider a typical loop integral $I(k)$ for a one-particle-irreducible (1PI) diagram contributing to $\Sigma$ in $(1+d)$-dimensional spacetime, with $L$ loops and $E$ internal edges,
\begin{equation}
I(k)=\int_p \Omega(k,p)\prod_{e=1}^{E} G_{e}^{(0)}(q_{e}),\label{eq:IntegralId}
\end{equation}
where $\int_{p}\equiv\int\prod_{l=1}^{L}\frac{d^{1+d}p_{l}}{(2\pi)^{1+d}}$, $k=(\omega,\mathbf{k})$ is the external momentum, and $p_l$ are independent loop momenta. Here $G_{e}^{(0)}(q_{e})$ denotes the $e$-th propagator with momentum $q_e$, while $\Omega(k,p)$ is a polynomial in $k$ and $p_l$. By energy-momentum conservation, each $q_e^\mu$ is a linear combination of $k^\mu$ and $p_l^\mu$. See Fig.~\ref{fig1} for an illustrative example.
Not every pole of $G_{e}^{(0)}(q_{e})$ gives rise to a singularity of $I(k)$. For fixed $\mathbf{k}$, a singularity of $I(k)$ occurs only when two poles approach the integration contour from opposite sides and pinch it as $\omega\to\omega_{(s)}$. 
This type of singularity is usually referred to as a Landau singularity.

Such Landau singularities are generally difficult to determine directly from the loop integral. The Landau equations \citep{Landau:1959fi,Eden:1966dnq} provide a systematic and simpler way to determine them:
\begin{subequations}
\begin{align}
\alpha_{e}D_e(q_{e})& =0,\quad e=1,\dots,E, \label{eq:LandauEq01}\\
\sum_{e}\alpha_{e}\frac{\partial }{\partial p_{l}^{\mu}}D_e(q_{e}) & =0, \quad l=1,\dots,L,\label{eq:LandauEq02}
\end{align}
\end{subequations}
where $D_e(q_{e})$ denotes the denominator of $G_{e}^{(0)}(q_{e})$, and and $\alpha_e$ are Feynman parameters satisfying $\alpha_e\in[0,1]$  
\footnote{We need not consider vanishing numerators at the Landau singularities separately. Since the numerator is polynomial in the momenta, the original loop integral can be decomposed into a linear combination of integrals whose numerators do not vanish at the solutions of the Landau equations. See also Ref.~\citep{Hannesdottir:2024hke} for a direct treatment of this case.}.
Solving Eqs.~(\ref{eq:LandauEq01}) and (\ref{eq:LandauEq02}), together with the constraint $\sum_{e}\alpha_{e}=1$, determines the allowed singularities. 

We first focus on the leading singularities, for which all $\alpha_{e}\neq0$. If some $\alpha_{e}=0$, the singularity is named non-leading and can be viewed as a leading singularity of a reduced diagram obtained by contracting the internal edges with $\alpha_{e}=0$ \citep{Landau:1959fi,Eden:1966dnq}. We will return to such non-leading singularities later.


\vspace{0.3cm}
\paragraph*{Self-energy and relevant topology.}--- We now analyze the leading singularities of the self-energy. The components of the self-energy matrix $\Sigma$ are not independent: they satisfy $\Sigma_{ra}(k)=\Sigma_{ar}(-k)$, while the fluctuation-dissipation theorem relates $\Sigma_{aa}$ to $\Sigma_{ar}$ \citep{Kamenev2011}. In addition, $\Sigma_{rr}$ vanishes because  $G_{aa}=\left\langle \varphi_{a}\varphi_{a}\right\rangle =0$. It is therefore sufficient to consider only the 1PI diagrams contributing to $\Sigma_{ar}$.

The simplest case is the two-vertex contribution, $V=2$, since diagrams with a single vertex can be regularized to zero \citep{Gao:2018bxz,Jain:2020zhu}. For $\Sigma_{ar}$, the only nonvanishing 1PI topology in this case is the banana diagram shown in Fig.~\ref{fig1}.
Using the standard identity for connected diagrams \citep{nakanishi1971graph},
$E-V+1=L$, the number
of internal edges is $E=L+1$. 
Since each interaction vertex in the effective Lagrangian (\ref{eq:EffectiveLagrangian}) involves at least one $\varphi_{a}$-leg, the banana diagram must contain a retarded propagator. 

Applying the pole decomposition illustrated in Eq.~(\ref{eq:GraGeneralForm0}) to both $G_R^{(0)}$ and $G_S^{(0)}$, and substituting the resulting forms into the loop integral (\ref{eq:IntegralId}), we express $I(k)$ as a sum of simpler terms,
$I(k)=\sum_{n_{1},...,n_{L+1}}\mathcal{I}(k;\{n_{e}\})$,  
where $n_e=1,\dots,N_0$ denotes the pole selected from the $e$-th propagator.  Each term takes the form
\begin{equation}
 \mathcal{I}(k;\{n_{e}\})\equiv\int_{p}\Omega^{\prime}(k,p)\prod_{e}(p^{0}_{e}-\mathfrak{w}_{n_e})^{-1},  
\end{equation}
where all irrelevant factors have been absorbed into the $\Omega^{\prime}.$ 
Here, $\mathfrak{w}_{n_e}$ denotes the location of the pole selected from the $e$-th propagator, with its dependence on the loop momentum $\mathbf{p}_e$ left implicit.
 Accordingly, $\mathcal{I}(k;\{n_e\})$ is labeled by the full set $ \{n_e\}$, which specifies one selected pole for each internal propagator.

\vspace{0.3cm}
\paragraph*{Leading singularities of banana diagrams.}---
We now derive the leading singularities of banana diagrams. Since hydrodynamic EFT concerns the low-momentum regime, we expand the basic modes at small momentum as
$\mathfrak{w}_{n_e}(\mathbf{p})=-i\gamma_{n_e}-c_{n_e}\left|\mathbf{p}\right|-i\Gamma_{n_e}\left|\mathbf{p}\right|^{2}+\mathcal{O}(\left|\mathbf{p}\right|^{3})$, where $\gamma_{n_e}$, $c_{n_e}$, and $\Gamma_{n_e}$ are constants satisfying $\mathrm{Re}\,\gamma_{n_e}\geq 0$, $c_{n_e}\in\mathbb{R}$, and $\Gamma_{n_e}>0$. These conditions ensure $\mathrm{Im}\,\mathfrak{w}_{n_e}<0$ for $\mathbf{p}\neq 0$.

Substituting the small-$\mathbf{p}$ expansion into the Landau equations (\ref{eq:LandauEq01}, \ref{eq:LandauEq02}), we obtain $\alpha_e=1/E$ for all $e$. We then find that the existence of a leading singularity in a given term $\mathcal{I}(k;\{n_e\})$ requires all selected $c_{n_e}$ to coincide with a common value $c$; otherwise,  leading singularities are  absent. Here, the value $c$ can depend on the chosen set $\{n_e\}$. We thus obtain the leading singularities of $\mathcal{I}(k;\{n_{e}\})$,
\begin{equation}
\omega_{(s)}=-i\gamma_{(s)}-c|\mathbf{k}|-i\Gamma_{(s)}|\mathbf{k}|^{2},\label{eq:D1Singulartity}
\end{equation}
where, for the chosen set $\{n_e\}$, $\gamma_{(s)}\equiv\sum_{e=1}^{E}\gamma_{n_e}$ and $\Gamma_{(s)}^{-1}\equiv\sum_{e=1}^{E}\Gamma_{n_e}^{-1}$. Different choices of $\{n_e\}$ generally lead to different singularities.
Eq.~(\ref{eq:D1Singulartity}) gives the general form of the leading singularities. Within the class of theories considered here,  its derivation does not depend on the loop order, the type of noise, or the detailed form of the nonlinear interactions. Some special cases of this result have appeared in Refs.~\citep{Delacretaz:2020nit,Grozdanov:2024fle} and references therein.

The singularities in Eq.~(\ref{eq:D1Singulartity}) correspond to infinitely many new modes generated by nonlinear interactions. Their damping is determined by the selected poles. In a given banana diagram, if some  $\gamma_{n_{e}}$ have nonzero real parts, then $\mathrm{Re}\,\gamma_{(s)}>\mathrm{Re}\,\gamma_{n_{e}}$, and the corresponding mode $\omega_{(s)}$ decays faster than the basic modes. It is therefore subleading at late times. By contrast, if all $\gamma_{n_e}$ are purely imaginary or vanish, the new mode can become important at late times, since $\Gamma_{(s)}$ is smaller than each of the $\Gamma_{n_e}$.


\vspace{0.3cm}
\paragraph{Singular behavior of self-energy.}---
We now turn to the singular nature of the self-energy. Eq.~(\ref{eq:D1Singulartity}) determines the locations of the leading singularities. The non-leading singularities are likewise included, since they can be identified with leading singularities of reduced diagrams obtained by contracting internal edges. Deriving the late-time behavior then requires the singular behavior in the vicinity of these points. We therefore consider a general Landau singularity.

Suppose that a singularity $\omega_{(s)}$ of $I(k)$ corresponds to a solution of the Landau equations with several $\alpha_e=0$ for $e=1,2,\dots,\kappa$. Keeping $\mathbf{k}$ fixed and expanding around this solution, with deviations denoted by $\delta\omega$, $\delta\alpha_e$, and $\delta p_e^\mu$, the singular part of $I(k)$ near $\omega_{(s)}$ arises from the integration region where $\delta\alpha_e$ and $\delta p_e^\mu$ are small. Following Refs.~\citep{Polkinghorne:1960udx,Hannesdottir:2024hke}, the dominant singular part takes the form
\begin{equation}
I(\omega_{(s)}+\delta\omega,\mathbf{k})\propto\begin{cases}
(\delta\omega)^{\sigma/2}\ln\delta\omega & \sigma=0,2,4,...,\\
(\delta\omega)^{\sigma/2} & \textrm{otherwise},
\end{cases}\label{eq:SingularPart}
\end{equation}
 with the integer $\sigma \equiv Ld-V+\kappa$.
This result shows that the singularities of $I(k)$ can be either poles or branch points. In particular, for spatial dimension $d=0$, $I(\omega)$ involves only frequency integration, and the singularities are poles. We also find that $\sigma\ge 0$ for $d\ge 2$, so the singularities of $I(k)$ are branch points in this case.

\vspace{0.3cm}
\paragraph*{Late-time behavior of correlation function.}---
We are now ready to derive the late-time behavior of the two-point correlation function $G$. Combining the poles of the leading-order propagator with the singularities of the self-energy, and carefully performing the inverse Fourier transform of Eq.~(\ref{eq:DysonEq}) using the singular behavior derived in Eq.~(\ref{eq:SingularPart}), we obtain
\begin{eqnarray}
    G(t,\mathbf{k}) & \approx &  \sum_{n=1}^{N_{0}}[C_{0}(\mathbf{k})+C_{1}(\mathbf{k})t]e^{-i\mathfrak{w}_{n}(\mathbf{k})t}\nonumber \\
 & &+\sum_{L,V,\kappa,\omega_{(s)}}\sum_{j=0}^{\infty}D(\mathbf{k})t^{-1-j-\sigma/2}e^{-i\omega_{(s)}(\mathbf{k})t},\label{eq:GSigma0}
\end{eqnarray}
where $C_{0}$ and $C_{1}$ depend on $n$ and $\mathbf{k}$, while $D(\mathbf{k})$ depends on $L$, $V$, $\kappa$, $\omega_{(s)}$, $j$, and $\mathbf{k}$. The first line of Eq.~(\ref{eq:GSigma0}) arises from the poles of $G^{(0)}$. The term proportional to $C_{1}(\mathbf{k})t$ appears because the second term on the right-hand side of Eq.~(\ref{eq:DysonEq}) contains two factors of $G^{(0)}$, so some poles become double poles. The second line arises from the singularities of $\Sigma$ through Eq.~(\ref{eq:SingularPart}). See the End Matter for the derivation.

If all basic modes $\mathfrak{w}_{n}=-i\gamma_n+\mathcal{O}(|\mathbf{k}|)$ are gapped and $\mathrm{Re}\,\gamma_{n}>0$ at $\mathbf{k}=0$ (for example, in Brownian motion), then all banana diagrams generate new modes of the form (\ref{eq:D1Singulartity}) that decay faster than the basic modes. In this case, the late-time behavior of $G(t,\mathbf{k})$ is controlled by the exponential factor $e^{-\gamma_{n}t}$ from the free propagator.

In most macroscopic systems, however, conservation laws imply the existence of gapless modes, so that $\gamma_n=0$ for some basic modes. As a result, Eq.~(\ref{eq:D1Singulartity}) can also yield modes with $\gamma_{(s)}=0$. For $\mathbf{k}\neq0$, these modes decay as $e^{-t\Gamma_{(s)}|\mathbf{k}|^{2}}$, which is slower than the basic modes and therefore dominates the late-time dynamics.

\vspace{0.3cm}
\paragraph*{Power-law late-time relaxation.}---
We now consider the long-wavelength limit of the gapless modes identified above and analyze the asymptotic behavior of $G(t,\mathbf{k}\to\mathbf{0})$. At first sight, the first line of Eq.~(\ref{eq:GSigma0}) appears to grow linearly with $t$, since for gapless basic modes the factor $e^{-i\mathfrak{w}_{n}t}$ becomes constant as $\mathbf{k}\to\mathbf{0}$. This linear term must be absent, as a nonzero $C_{1}(\mathbf{0})$ would signal an instability. Hence, in the $\mathbf{k}\to\mathbf{0}$ limit, the first line of Eq.~(\ref{eq:GSigma0}) generally reduces to a constant plus exponentially decaying terms.
In contrast, the second line of Eq.~(\ref{eq:GSigma0}) exhibits power-law decay in the limit $\mathbf{k}\to\mathbf{0}$. The slowest decay follows from $j=0$, $\kappa=0$, and $L=L_{\textrm{min}}$, with $L_{\min}$ the minimum number of loops. Restricting further to the lowest perturbative order, the dominant contribution comes from the banana diagram in Fig. \ref{fig1}.
We therefore obtain
\begin{equation}
\lim_{\mathbf{k}\rightarrow\mathbf{0}}G(t,\mathbf{k})\approx\mathrm{const}+D(\mathbf{0})t^{-L_{\textrm{min}}d/2}.\label{eq:LongTimeTails}
\end{equation}
The power-law factor $t^{-L_{\min}d/2}$ reflects the slow late-time relaxation generated by long-range correlations in space and time.
It arises not from the gapless modes of the linearized theory alone, but from nonlinear mode coupling encoded in the self-energy $\Sigma$. In the absence of nonlinearity, $\Sigma$ vanishes, and the power-law behavior disappears.

Remarkably, the power-law behavior $t^{-L_{\min}d/2}$ is universal in the sense that it is determined only by the spatial dimension $d$ and the interaction structure, with the latter entering through $L_{\min}$.
$L_{\min}$ denotes the smallest loop number for which a nonvanishing banana diagram exists. Equivalently, it is determined by the lowest nonlinear power $\varphi_r^{1+L_{\min}}$ in the equation of motion, corresponding to a vertex $\varphi_a\varphi_r^{1+L_{\min}}$ in $\mathcal{L}_{\mathrm{int}}$.
For example, if the cubic interaction $\varphi_a\varphi_r^2$ is present in Eq.~(\ref{eq:EffectiveLagrangian}), then the one-loop banana diagram contributes and $L_{\min}=1$, as in Refs.~\citep{Chen-Lin:2018kfl,Abbasi:2022aao,Jain:2020zhu,Grozdanov:2024fle}. If $\varphi_a\varphi_r^2$ is absent, then $L_{\min}\ge 2$, since no nonvanishing one-loop banana diagram can be constructed.

We thus arrive at a general picture of late-time relaxation and the emergence of power-law decay from nonlinear mode coupling. We notice that in earlier studies \citep{Arnold:1997gh,Kovtun:2003vj}, such power-law decay was analyzed for the two-point correlation function of the charge current $\mathbf{j}$, whose nonlinear part can be written simply as the product of the number density $\rho$ and the velocity $\mathbf{v}$. Under the Gaussian approximation, the correlator $\langle\mathbf{j}\mathbf{j}\rangle$ factorizes into $\langle\rho\rho\rangle\langle\mathbf{v}\mathbf{v}\rangle$, corresponding in the SK-EFT language to a one-loop diagram built from two propagators. Their derivation is therefore equivalent to evaluating a one-loop integral and yields Eq.~(\ref{eq:LongTimeTails}) with $L_{\min}=1$. This strategy was recently extended in Ref.~\citep{Delacretaz:2020nit} to correlations of operators with more indices. Our approach provides a complementary route that does not require an explicit construction of the nonlinear part of the operator and thus applies more directly when the quantity of interest obeys a nonlinear equation that is difficult to solve. 




\vspace{0.3cm}
\paragraph*{Summary.}---
In this work, we use Landau singularity analysis to determine the general late-time behavior of two-point correlation functions in the SK-EFT. This approach bypasses explicit loop integration and instead identifies new decay modes directly from algebraic equations. Starting from a generic effective Lagrangian that incorporates both classical dynamics and stochastic fluctuations, we analyze the structure of two-point functions and clarify how nonlinear interactions enter the late-time dynamics through loop corrections. Using the Landau equations, 
we obtain the general form of the leading singularities in banana diagrams and show that nonlinear interactions generate infinitely many new decay modes. We further determine the singular part of the loop integrals near the Landau singularities in frequency space and, by transforming to the time domain, derive the corresponding late-time behavior. In particular, when gapless modes are present, the late-time dynamics is governed by the new modes induced by nonlinear interactions, producing power-law decay. 

Our results provide a unified description of late-time behavior in a broad class of systems and clarify the link between Landau singularities in frequency space and  power-law late-time relaxation. They are also useful for analyzing higher-loop diagrams and non-Gaussian noise effects in correlation functions. 


\vspace{0.5cm}
\paragraph{\bf Acknowledgments.}
This work was supported in part by National Key Research and Development Program of China under Contract No. 2022YFA1605500, Chinese Academy of Sciences (CAS) under Grant No. YSBR-088, and National Natural Science Foundation of China (NSFC) under Grant No.~12135011 and 125B2110.

\bibliographystyle{apsrev4-2}
\bibliography{Refs}

\setcounter{equation}{0}
\renewcommand{\theequation}{A\arabic{equation}}
\onecolumngrid
\section*{End Matter}
\twocolumngrid
\paragraph*{Inverse Fourier transformation in frequency.}---
To obtain Eq.~(\ref{eq:GSigma0}), we take the inverse Fourier transform of Eq.~(\ref{eq:DysonEq}) to the time domain. The first line of Eq.~(\ref{eq:GSigma0}) follows directly from the poles of the free propagator $G^{(0)}$ by the residue theorem. We now turn to the second line, which is determined by the singularities of $I(\omega)$. For simplicity, we suppress the dependence on $\mathbf{k}$.
Near a singularity $\omega_{(s)}$ in the lower half-plane, we write
\begin{equation}
I(\omega)=F_{(s)}(\omega)+F_{\mathrm{rest}}(\omega),\label{eq:Idecomp}
\end{equation}
where $F_{\mathrm{rest}}(\omega)$ is regular at $\omega_{(s)}$, while $F_{(s)}(\omega)$ has the form
$F_{(s)}(\omega)=\phi_a(\omega)(\omega-\omega_{(s)})^{\sigma/2}X_a(\omega),
$ with $X_1(\omega)=\ln(\omega-\omega_{(s)})$ for $\sigma=0,2,4,\dots$, and $X_2(\omega)=1$ otherwise. Here $\phi_{1,2}(\omega)$ is analytic in the lower half-plane. The contribution of $\omega_{(s)}$ to the inverse transform is therefore
\begin{equation}
\widetilde{F}(t)=\int_{-\infty}^{+\infty}\frac{d\omega}{2\pi}\,F_{(s)}(\omega)e^{-i\omega t},
\end{equation}
where $t>0$. We assume that $\phi_{1,2}(\omega)$ vanishes sufficiently fast at infinity so that the integral converges.

Let us first consider the case $\sigma=0,2,4,\dots$. In the lower half-plane, the only singularity is the branch point at $\omega_{(s)}$ associated with $\ln(\omega-\omega_{(s)})$. Deforming the contour around the branch cut and using the residue theorem, we obtain 
\begin{eqnarray}
\widetilde{F}(t) & = & \mathcal{A}(t)
\left(\int_{+\epsilon} -\int_{-\epsilon} \right)
\frac{dz}{2\pi}\phi_{1}(z+\omega_{(s)}) \mathcal{B}(z,t)\ln z, \nonumber \\
\label{eq:integral01}
\end{eqnarray}
where $\mathcal{A}(t)\equiv e^{-i\omega_{(s)}t}$, $\mathcal{B}(z,t)=e^{-izt}z^{\frac{\sigma}{2}}$, $z\equiv\omega-\omega_{(s)}$, and the contours $\pm\epsilon$ run from $\pm\epsilon$ to $-i\infty\pm\epsilon$, with $\epsilon>0$ infinitesimal. Along the $+\epsilon$ contour, $\arg z=-\pi/2$, while along the $-\epsilon$ contour, $\arg z=3\pi/2$. Eq.~(\ref{eq:integral01}) then reduces to
\begin{eqnarray}
\widetilde{F}(t)
 & = & -(-i)^{\frac{\sigma}{2}}\mathcal{A}(t)\int_{0}^{+\infty}dz\mathcal{B}(z,-it)\phi_{1}(\omega_{(s)}-iz).\label{eq:Fexpr01} \nonumber \\
\end{eqnarray}
As $t\rightarrow+\infty$, we can use the Watson's lemma \citep{miller2006applied} in asymptotic
analysis to perform the integral in Eq.~(\ref{eq:Fexpr01}):
\begin{equation}
\widetilde{F}(t)=-\mathcal{A}(t)\sum_{n=0}^{\infty}(-i)^{n+\frac{\sigma}{2}}\mathcal{C}(\sigma)\phi_{1}^{(n)}(\omega_{(s)})t^{-1-n-\frac{\sigma}{2}},\label{eq:Fexpr02}
\end{equation}
where $\mathcal{C}(\sigma)\equiv \Gamma\left(1+n+\frac{\sigma}{2}\right)/n!$.
The equality in Eq.~(\ref{eq:Fexpr02}) should be understood as an asymptotic
relation. 

When $\sigma=\pm1,\pm3,\dots$, the singularity of $F_{(s)}(\omega)$ at $\omega_{(s)}$ is a branch point associated with $(\omega-\omega_{(s)})^{\sigma/2}$. Repeating the derivation leading to Eq.~(\ref{eq:Fexpr02}), we obtain the large-$t$ asymptotic form
\begin{equation}
\widetilde{F}(t)=\frac{\mathcal{A}(t)}{\pi}\sum_{n=0}^{\infty}(-i)^{1+n+\frac{1}{2}\sigma}\mathcal{C}(\sigma)\phi_{2}^{(n)}(\omega_{(s)})t^{-1-n-\frac{\sigma}{2}}.\label{eq:Fexpr03}
\end{equation}

The remaining cases are $\sigma=-2,-4,-6,\dots$, for which the singularity of $F_{(s)}(\omega)$ at $\omega_{(s)}$ is a pole. Applying the residue theorem, we find  
\begin{eqnarray}
    \widetilde{F}(t)&=&\frac{\mathcal{A}(t)}{(-1-\frac{\sigma}{2})!}\sum_{n=0}^{-\sigma/2-1}\left(\begin{array}{c}
-1-\frac{\sigma}{2}\\
n
\end{array}\right)(-i)^{-n-\frac{\sigma}{2}}
\nonumber \\ && \times
\phi_{2}^{(n)}(\omega_{(s)})t^{-1-n-\frac{\sigma}{2}}.\label{eq:Fexpr04}
\end{eqnarray}

Finally, combing Eqs. (\ref{eq:Fexpr02} - \ref{eq:Fexpr04}), we
obtain the second line of Eq.~(\ref{eq:GSigma0}).

\end{document}